\documentclass[aoas,MSNbibl,nameyear,dvips]{arximspdf}
\usepackage{dcolumn}
\usepackage{graphicx}

\doi{10.1214/11-AOAS516}
\volume{6}
\issue{2}
\pubyear{2012}
\firstpage{432}
\lastpage{456}

\makeatletter

\let\sv@tabnotetext\tabnotetext
\let\sv@tabnotemark@fmt\tabnotemark@fmt
\long\def\legend#1{{\let\tabnote@indent\leavevmode\sv@tabnotetext[]{}{#1}}}

\newcolumntype{d}[1]{D{.}{.}{#1}}

\newcommand{\veb}[1]{\bolds{#1}}
\newcommand{\vef}[1]{\mathbf{#1}}

\newtheorem{prop}{Proposition}

\newtheorem{lemma}{Lemma}

\makeatother

\begin{document}
\begin{frontmatter}

\title{Simultaneous SNP identification in association studies with missing data}
\runtitle{Simultaneous SNP identification}

\begin{aug}
\author[A]{\fnms{Zhen} \snm{Li}},
\author[B]{\fnms{Vikneswaran} \snm{Gopal}},
\author[C]{\fnms{Xiaobo} \snm{Li}},
\author[C]{\fnms{John~M.}~\snm{Davis}}
\and
\author[D]{\fnms{George} \snm{Casella}\corref{}\ead[label=e1]{casella@stat.ufl.edu}}
\runauthor{Z. Li et al.}
\affiliation{State Street Corporation, University of Florida,
University of Florida, University~of~Florida and University of Florida}
\address[A]{Z. Li\\
State Street Corporation\\
1 Lincoln Street, 15th floor\\
Boston, Massachusetts 02111\\
USA} %adresu isvedimo komanda gale!
\address[B]{V. Gopal\\
Department of Statistics\\
University of Florida\\
Gainesville, Florida 32611\\
USA}
\address[C]{X. Li\\
J. M. Davis\\
School of Forest Resources\\
\quad and Conservation\\
University of Florida\\
Gainesville, Florida 32611\\
USA}
\address[D]{G. Casella\\
Department of Statistics\\
\quad and Genetics Institute\\
University of Florida\\
Gainesville, Florida 32611\\
USA}
\end{aug}

% HISTORY:
\received{\smonth{12} \syear{2010}}
\revised{\smonth{9} \syear{2011}}

% ABSTRACT
%
\begin{abstract}
Association testing aims to discover the underlying relationship
between genotypes (usually Single Nucleotide Polymorphisms, or\break SNPs)
and phenotypes (attributes, or traits). The typically large data sets
used in association testing often contain missing values. Standard
statistical methods either impute the missing values using relatively
simple assumptions, or delete them, or both, which can generate biased
results. Here we describe the Bayesian hierarchical model BAMD
(Bayesian Association with Missing Data). BAMD is a Gibbs sampler, in
which missing values are multiply imputed based upon all of the
available information in the data set. We estimate the parameters and
prove that updating one SNP at each iteration preserves the ergodic
property of the Markov chain, and at the same time improves
computational speed. We also implement a model selection option in
BAMD, which enables potential detection of SNP interactions.
Simulations show that unbiased estimates of SNP effects are recovered
with missing genotype data. Also, we validate associations between
SNPs and a carbon isotope discrimination phenotype that were previously
reported using a family based method, and discover an additional SNP
associated with the trait.
%We conclude that BAMD provides biologists with an objective tool to
%explore possible epistatic interactions among genes that condition a
%phenotype, and to explore the basis of genetic correlation among
%phenotypes.
BAMD is available as an \texttt{R}-package from
\url{http://cran.r-project.org/package=BAMD}.
\end{abstract}

% KEYWORDS
%

\end{frontmatter}

%s1 #&#
\section{Introduction}

This work was motivated from a study of the genomics of loblolly pine,
an economically and ecologically important tree species in the United
States. The native range of loblolly pine extends from Maryland, south
to Florida, and west to Texas. Its annual harvest value is
approximately $19$ billion dollars [\citet{McKHow96}]. The
pine species in the southern states produces $58\%$ of the timber in
the US and $15.8\%$ of the world's timber [\citet{WeaGre02}]. We
are interested in discovering the relationship between phenotypic
traits and genes underlying complex traits in loblolly pine, so we can
understand their evolution and apply that knowledge to genetic
improvement. We are especially interested in SNPs associated with
disease resistance and response to water deficit.

Large genomic data sets typically contain missing data. Missing data
create imbalance and complicate calculations required for statistical
analyses. There are various approaches to dealing with missing data.
Eliminating cases is one approach, but undesirable in large data sets
where most or all cases have missing data. Imputation is more commonly
used [\citet{Hui00}, \citet{Daietal06}]. Single imputation using
haplotype data [\citet{Maretal07}, \citet{Suetal}, \citet{SunKar08},
\citet{SzaBea}], either implicitly or explicitly,
relies on linkage disequilibrium among markers, or information that can
be extracted from other data sets [\citet{SteSmiDon01},
\citet{SchSte06}, \citet{SerSte07}]. However, there is no
reference genome sequence for loblolly pine, so it is not possible to
impute missing SNPs from flanking SNPs.

It is well established that single imputation approaches, while fast,
can give biased parameter estimates [\citet{GreFin95}; see also
\citet{Heietal06}]. The best approach is to average over the
missing data using the formal missing data distribution, rather than to
impute a single value based on a possibly ad hoc scheme. This is
appealing because it addresses uncertainty and variability in the
missing data [\citet{Little02},
\citet{Daietal06}], particularly in species or genomic regions
where LD decays rapidly and thus adjacent SNPs are not necessarily
correlated [\citet{autokey7}, \citet{NeaIng08}]. However,
multiple imputation is so computationally intensive that, prior to the
present work, it has not been feasible for larger genomic data sets.

Several approaches have been developed recently to enable association
testing. Association testing identifies relationships between
polymorphisms in DNA sequence (most commonly Single Nucleotide
Polymorphisms, or SNPs) and phenotypes, as a strategy to identify the
genes that control traits [\citet{autokey7},
\citet{HirDal05}, \citet{Bal06}]. For family-based analysis,
\citet{CheAbe07} used an identity by descent parameter to measure
correlation among SNPs, and a kinship coefficient to model the
correlation among siblings to develop the Quantitative Transmission
Disequilibrium Test (QTDT). Other approaches allow association testing
in populations with recent mating or historical (unrecorded) mating, or
combinations. TASSEL fits a mixed model to detect associations while
taking into account both ``coarse-scale'' relatedness based on
population structure [\citet{PriSteDon00}] and ``fine-scale''
relatedness based on a matrix of kinship coefficients
[\citet{Yuetal06}]. Our approach for family based analyses
accomplishes the same goal by employing the numerator relationship\vadjust{\goodbreak}
matrix [\citet{autokey12}; see also \citet{Quaas}], which
avoids complications arising from nonpositive definite matrices derived
from complex interrelationships. A desirable feature of any association
testing approach is simultaneous solution of multiple SNP effects to
prevent upward bias in parameter estimates, and to appropriately model
the underlying biological system in which many SNPs act in concert to
condition the phenotype. Such an approach is developed in
\citet{Wiletal10}, who introduce Multilevel Inference for SNP
Association (MISA), using imputation from fastPHASE
[\citet{SteSmiDon01}, \citet{SerSte07}] and a
Bayes-driven stochastic search method to find good models.

In this paper we introduce BAMD (Bayesian Association with Missing
Data) and show that computation time required for formal multiple
imputation can be reduced without sacrificing accuracy, establishing
the feasibility of using BAMD on genomic data sets. Our approach is to
use all available data in imputation of missing SNPs. This approach is
motivated statistically; we use all of the available information to
estimate SNP effects on phenotypes, across all possible values for
missing SNPs. Prior knowledge such as pedigree structure may be used as
constraints. Simulations show that BAMD detects SNP effects efficiently.

On the loblolly pine genomic data, we used a series of \textit{tag SNPs}
[\citet{Gonetal08}]. Tag SNPs are markers that are relatively
evenly dispersed throughout the genome, and are used to survey
chromosomal segments for genes that underly phenotypes. One assessment
of the performance of BAMD was to use it on this same population,
genotype, and phenotype data, in which it had been found that three of
the tag SNPs were significantly associated with carbon isotope
discrimination [a measure of water use efficiency,
\citet{Gonetal08}]. BAMD detected a fifth tag SNP in addition to
the other four tag SNPs that were detected using QTDT in that previous
work.

An additional feature of BAMD is the variable selector. The variable
selector searches model space for the most parsimonious set of SNPs
that explain the phenotype. This feature is designed for unsupervised
discovery of interactions among SNPs, and should find application in
situations where epistatic interactions are important determinants of phenotype.

The remainder of the paper is organized as follows. In Section \ref
{secmodel} we describe the model and the estimation of parameters, and
Section \ref{secselect} describes the variable selector, including the
use of Bayes factors, the stochastic search, and computational
strategies. In Section \ref{subsecmissingness}, we investigate the
amount of missing data that BAMD can handle through a simulation.
Section \ref{subseccompare} compares our procedure to BIMBAM, a popular
genomics program that does both imputation and variable selection.
Section \ref{secanalysis} analyzes the loblolly pine data, where we
discover a previously undiscovered SNP. Section \ref{secdisc} contains
a concluding discussion. Computational implementation is
described in the \hyperref[app]{Appendix}, and the
accompanying theorems and proofs can be found in the
online Supplemental Information [\citet{Lietal}].\vspace*{-2pt}

%s2 #&#
\section{Model}\label{secmodel}

Our method can be viewed as a two-stage procedure. The first stage involves
identifying individual SNPs that have significant effects on the phenotype,
with all SNPs in the model. The second stage searches for the best subset
of SNPs, from those picked out in the first stage. First we describe
the model.\vspace*{-2pt}

%s2.1 #&#
\subsection{Conceptual framework for BAMD}
The response is assumed to be continuous, following a normal
distribution. The
data set has fully observed family covariates for all the
observations. Missing values are imputed only among SNPs,
although the method can be modified to impute missing values for
phenotypes as
well. We focus on testing the relationship between the response
and the SNPs. We assume only additive effects among SNPs, although the
method can
be adapted to quantifying additive and dominance effects
of SNPs.

%We use a variation of the linear mixed model of Yu {\it\etal} (2006),
%given by
We begin with the linear mixed model
%
%e1 #&#
%
\begin{equation}
\label{eqfixedeffects}
\vef{Y} = \vef{X} \veb{\beta} + \vef{Z} \veb{\gamma} + \veb
{\varepsilon},
\end{equation}
where $\vef{Y}_{n \times1}$ is the phenotypic trait, $\vef{X}_{n
\times
p}$ is the design matrix for family covariates, $\veb{\beta}_{p \times
1}$ are the coefficients for the family effect, $\vef{Z}_{n \times s}$
is the design matrix for SNPs (genotypes), $\veb{\gamma}_{s \times1} $
are the coefficients of the additive effect for SNPs, and $\veb
{\varepsilon}_{n \times1} \sim N(0,\sigma^2 \vef{R})$. The matrix
$\vef
{R}$ is the numerator relationship matrix,
describing the degree of kinship between different individuals.
(Details on the calculation of $\vef{R}$ are given in Appendix \ref
{appnumerator}.)

For our application here (carbon isotope data), we have $n=1000$ and
$s=450$. In another application of BAMD [\citet{QueGopCum}], they
used $n=450$ and $s=400$. In both cases the number of covariates, $p$,
was less than~$6$. With a fully Bayesian implementation, it is possible
to adapt BAMD to the $p \gg n $ case.

Each row of the matrix $\vef{Z}$, $\vef{Z}_i, i=1, \ldots, n,$ corresponds
to the
SNP genotype information of one individual, which can be homozygous for
either of the two nucleotides $(-1,1)$ or heterozygous $(0)$. Some of
this information may be
missing, and we write $\vef{Z}_i = (\vef{Z}_i^{\mathrm{obs}}, \vef{Z}_i^{
\mathrm{miss}})$, where $\vef{Z}_i^{\mathrm{obs}}$ are the observed genotypes for the
$i$th individual, and
$\vef{Z}_i^{\mathrm{miss}}$ are the missing genotypes. Note two aspects of
this framework:
\begin{longlist}[(2)]
\item[(1)] The values of $\vef{Z}_i^{\mathrm{miss}}$ are not observed. Thus, if
$*$ denotes one
missing SNP, a possible $\vef{Z}_i$ is
$
\vef{Z}_i = (1,*,0,0,*,*,1)
$.
\item[(2)] Individuals are likely to have missing data at different SNP loci.
So for 2 different individuals, we might have
\[
\vef{Z}_i = (1,*,0,0,*,*,1) \quad\mbox{and}\quad
\vef{Z}_{i^\prime} = (*,*,1,0,0,1,1).
\]
\end{longlist}

For a Bayesian model, we want to put prior distributions on the parameters.
We put a noninformative uniform\vadjust{\goodbreak} prior for $\veb{\beta}$, which essentially
leads us to least squares estimation. For $\veb{\gamma}$, we use the normal
prior $\veb{\gamma} \sim N(\vef{0}, \sigma^2\phi^2 \vef{I}_s)$. Here
$\phi^2$
is a scale parameter for the variance and $\sigma^2$ is the variance
parameter. For $\sigma^2$ and~$\phi^2$, we use inverted Gamma priors:
$\sigma^2 \sim \operatorname{IG}(a,b)$ and $\phi^2 \sim \operatorname{IG}(c,d)$,
where IG stands
for the
inverted Gamma distribution, and~$a, b, c,$ and $d$ are constants used
in the
priors. %We say $\sigma^2$ has $IG(a,b)$ distribution if
%$$f(\sigma^2)=\frac{b^a}{\Gamma(a)}\frac{\exp(-\frac{b}{\sigma^2})}{(
%{a+1}}$$
For specified $a, b, c, d$, the resulting posterior distribution is proper
[see \citet{HobCas96}].

We consider the case of tag SNP markers in the loblolly pine genome
with no significant linkage disequilibrium between them
[\citet{Gonetal06}]. Therefore, noninformative priors are
used for the missing SNPs, meaning that missing data have equal
probability of any allelic state. As information increases due to
higher marker density, or parental information, or allele frequency in
the population, missing data imputation could be constrained
accordingly.

We assume that missing SNPs in the data set are \textit{Missing at
Random} (\textit{MAR}). %In
%other words, we assume that the probability of whether the SNP is
%missing is
%related to the observed data, such as the phenotypic trait or other
%observed
%SNPs, and is independent of unobserved information.
In particular, let the value of the random variable $T$ denote whether~$Z$ is observed, with $T=1$ if the value is observed and $T=0$ if it is
missing. %Denote the observed part of $Z$ by $Z_{obs}$, the missing
%part by $Z_{miss}$, then $Z=(Z_{obs},Z_{miss})$.
If $\xi$ is the parameter of the missing mechanism, then, under the
model (\ref{eqfixedeffects}), the MAR assumption results in
\[
P(T|Y,Z^{\mathrm{obs}},Z^{\mathrm{miss}},\xi)=P(T|Y,Z^{\mathrm{obs}},\xi).
\]
So the distribution of the missing SNP could depend on the observed
SNPs, and the observed phenotypes. Of course, this does not require
such a dependence, it only allows for it. %In the absence of other
%information, or stronger assumptions, we cannot relax this assumption
%about the missing data mechanism.

Other assumptions about missing data mechanisms are less common than
MAR. The strongest assumption, and the most difficult to justify, is
\textit{Missing Completely at Random} (\textit{MCAR}). Under this assumption, the
missing data distribution is independent of all observed data, the
complete cases can be regarded as sub-samples from the population, and
statistical inference with regard to the complete cases is totally
valid. Under the model (\ref{eqfixedeffects}), the MCAR assumption can
be expressed as
\[
P(T|Y,Z^{\mathrm{obs}},Z^{\mathrm{miss}},\xi)=P(T|\xi).
\]
MCAR is regarded as unrealistic and, in most cases, it is not
satisfied. It is typically not used to model missing data, and we do
not use it here.

Conditional on the MAR assumption, we impute the missing SNPs based on
the correlation between SNPs
within individuals and between individuals, and use the phenotypic trait
information to improve the power of imputation. %MAR is a reasonable
%assumption
%and not as strict as missing completely at random, MCAR.

In this model, the covariance matrix, $\vef{R}$, models the covariance between
individuals within the same family, and covariance between individuals across
families. Phenotypic traits of related individuals are alike because they
share some proportion of SNPs, and genotypes of relatives are similar because
they share the same alleles passed on from parents. Various methods can be
used to calculate\vadjust{\goodbreak} the relationship matrix, such as using a co-ancestry
matrix, a~kinship matrix, etc. The basic idea is to calculate the probability
that 2 individuals share SNPs that are identical by descent. Some
methods use
pairwise calculations and thus do not guarantee a positive definite
relationship matrix, which is unsatisfactory when the relationship matrix
is used as covariance matrix. We use the recursive calculation method of
\citet{autokey12}, which gives a numerator relationship matrix that quantifies
the probability of sharing a SNP from the same ancestry, based on known
family pedigree and parent pedigree in the population. So by calculating
this relationship matrix we obtain a probability of 0.5 for the case
that two
siblings are within the same control-pollinated family and therefore
share the
same copy of a SNP, or a $0.25$ probability if the two siblings only
have one
parent in common. For the complex pedigree that we analyze here, there
are a total of $9$
categories of relatedness. %, with some additional details given in
%Supplemental Information \ref{appnumerator}.

%s2.2 #&#
\subsection{Estimation of parameters}\label{secgibbs}

The model (\ref{eqfixedeffects}) along with the prior specification
allows the use of a Gibbs sampler to estimate parameters. We can
iteratively sample from the the full conditionals, given by
%
%e2 #&#
%
\begin{eqnarray}\label{eqGibbs1}
\beta&\sim& N\bigl((X^\prime R^{-1} X)^{-1} X^\prime R^{-1}
(Y-\vef{Z}\gamma),
\sigma^2 (X^\prime R^{-1} X)^{-1} \bigr),\nonumber\hspace*{-30pt}\\
\gamma&\sim& N\biggl(\biggl(\vef{Z}^\prime R^{-1}
\vef{Z}+\frac{I}{\phi^2}\biggr)^{-1} \vef{Z}^\prime R^{-1} (Y-X\beta),
\sigma^2
\biggl(\vef{Z}^\prime R^{-1} \vef{Z}+ \frac{I}{\phi^2}\biggr)^{-1}
\biggr),\nonumber\hspace*{-30pt}\\
\sigma^2 &\sim& \frac{1}{(\sigma^2)^{n/2+s/2+a+1}}\hspace*{-30pt}\\
&&{}\times \exp
\biggl(-\frac{(Y-X\beta-\vef{Z}\gamma)^\prime
R^{-1}(Y-X\beta-\vef{Z}\gamma)+{|\gamma|^2}/{\phi
^2}+2b}{2\sigma^2}
\biggr),\nonumber\hspace*{-30pt}\\
\phi^2 &\sim& \frac{1}{(\phi^2)^{{s}/{2}+c+1}}\exp
{-\frac{({|\gamma|^2}/{\sigma^2}+2d)}{2\phi^2}}
.\nonumber\hspace*{-30pt}
\end{eqnarray}

The SNPs are contained in the $\vef{Z}$ matrix, which includes both the
observed SNPs and
missing SNPs, and we use the Gibbs sampler to impute the missing SNPs.
The Gibbs sampler
for the missing data simulates the samples of $Z_i^m$ according to the
distribution of
each missing SNP conditional on the rest of observed SNPs and sampled
missing SNPs. For
a particular SNP $Z^m_{ij}$, the $j$th missing SNP in the $i$th
individual, the
conditional distribution given the rest of the vector $Z^m_{i(-j)}$ and
all other
parameters in the model is
%
%e3 #&#
%
\begin{eqnarray}\label{eqconditional}
&&P\bigl(Z^m_{ij}=c | Z^m_{i(-j)} \bigr)\nonumber\hspace*{-30pt}\\[-8pt]\\[-8pt]
&&\qquad=\frac{\exp( - (Y_i-X_i\beta-Z_i^o \gamma_i^o- Z^m_{i(-j)}\gamma
^m_{i(-j)} - c\gamma^m_{ij})^2/(2\sigma^2))}{\sum_{\ell=1}^3\exp( -
(Y_i-X_i\beta-Z_i^o \gamma_i^o-
Z^m_{i(-j)}\gamma^m_{i(-j)} - c_\ell\gamma^m_{ij})^2/(2
\sigma^2))}.\hspace*{-30pt}\nonumber
\end{eqnarray}
The value $c$ is the genotype currently being considered for that
missing SNP, and $c_l$
represents any one of the possible\vadjust{\goodbreak} genotypes for the SNP. Notice there
are only $3$
terms in the denominator sum for each SNP and this is a key point why
Gibbs sampling
is computationally feasible for our situation with many SNPs and many
observations. We also note that the EM algorithm, which provides an
alternative method of parameter estimation, can require a prohibitive
amount of computation. See Appendix \ref{appEM}.

%s3 #&#
\section{Variable selection}\label{secselect}

The Gibbs sampler will estimate the full set of parameters in model
(\ref{eqfixedeffects}). However, it is often the case that a small set
of SNPs will explain a sufficient proportion of the variability that
might also be biologically meaningful. To this end, along with the
Gibbs sampler, we run a second Markov chain that searches through the
space of available models, looking for the one with the largest Bayes factor.

A model is specified by a vector $\delta$ of length $s$, whose entries
are either~$0$ or $1$. The $\gamma$ vector of model (\ref
{eqfixedeffects}) becomes $\gamma_\delta= \gamma\star\delta$, where
``$\star$'' denotes componentwise multiplication. The corresponding
columns of $\vef{Z}$ are deleted, giving $\vef{Z}_\delta$, and the
reduced model is
%
%e4 #&#
%
\begin{equation}
\label{eqreduced}
\vef{Y} = \vef{X} \veb{\beta} + \vef{Z}_\delta \veb{\gamma}_\delta
+ \veb
{\varepsilon}.
\end{equation}
Thus, the components of $\gamma_i$ corresponding to $\delta_i=0$ are
excluded from the model. Correspondingly, let $\veb{\theta}$ denote the
random vector consisting of all parameters in the full model, so $\veb
{\theta} := (\veb{\beta}, \veb{\gamma}, \sigma^2, \phi^2, \vef{Z} )$
and, naturally, $\veb{\theta}_\delta:= (\veb{\beta},
\veb{\gamma}_\delta, \sigma ^2, \phi^2, \vef{Z}_\delta)$. Let
$m_\delta, \pi_\delta$, and $p_\delta$ denote the marginal distribution
of $\vef{Y}$, the prior distribution on~$\veb{\theta}_{\delta}$, and
the conditional distribution of~$\vef{Y}$, respectively. We also write,
if needed, $\veb{\theta}=(\veb{\theta }_\delta ,
\veb{\theta}_{\delta^c})$, the latter containing the remaining
parameters not specified by $\delta$. For the full model containing all
parameters we omit the subscript.

%s3.1 #&#
\subsection{Searching with Bayes factors}

In order to compare models, we shall use the Bayes factor comparing
each candidate model to the full model, given by
%
%e5 #&#
%
\begin{equation}
\label{eqBFdef}
\mathrm{BF}_\delta=
\frac{m_\delta(\vef{Y})}{m(\vef{Y})} =
\frac{\int\pi_\delta(\veb{\theta}_\delta) p_\delta(\vef{Y}|\veb
{\theta
}_\delta)\,
\mathrm{d}\veb{\theta}_{\delta}}
{\int\pi(\veb{\theta}) p(\vef{Y}|\veb{\theta}) \,\mathrm{d}\veb
{\theta}},
\end{equation}
where $p$ denotes the full model. We now can compare models $\delta$
and $\delta^\prime$ through their Bayes factors, as a larger Bayes
factor corresponds to a model that explains more variability, when
compared to the full model. These pairwise comparisons result in a
consistent model selector [\citet{OHaFor04}], and have an
advantage over BIC, which is overly biased toward smaller models
[\citet{Casetal09}].

We now set up a Metropolis--Hastings (MH) search that has target
distribution proportional to the the Bayes factor, $\mathrm{BF}_\delta$. Given
that we are at model~$\delta$, we choose a candidate $\delta^\prime$
from a random walk (choose one component at random and switch $0
\rightarrow1$ or $1 \rightarrow0$) with probability $a$ and, with
probability $1-a$, we do an independent jump. This is a symmetric
candidate, and $\delta^\prime$ is accepted with probability $\min\{1,
\mathrm{BF}_{\delta^\prime}/\mathrm{BF}_{\delta}\}$.

%We do this by drawing samples from the model space according
%to the distribution $B(\delta)$, where $B(\delta) \propto\mathrm{BF}_
%For $\delta_1, \delta_2$, we let
%$d(\delta_1, \delta_2) = s - \delta_1^\prime\delta_2$ be the number of
%co-ordinates that differ between $\delta_1$ and $\delta_2$.
%We sample using
%a Metropolis--Hastings algorithm, with a candidate that is a
%mixture of a random walk and a uniform distribution on the sample
%space.
%It is a symmetric candidate, because
%q(\delta_2 | \delta_1) = p \cdot I\{d(\delta_1, \delta_2) = 1\}
%+ (1 - p) \cdot\frac{1}{2^s}
%= q(\delta_1 | \delta_2)
%and the acceptance probability for this M--H chain is
%= \min[ \frac{\mathrm{BF}_{\delta_2}}
%{\mathrm{BF}_{\delta_1}} , 1]

%s3.2 #&#
\subsection{Estimating the Bayes factor}

Calculating the Bayes factor in (\ref{eqBFdef}) requires knowing the
$\vef{Z}$ matrix, which is not the case with missing data. Thus, to
calculate the Bayes factor, we need to use the imputed $\vef{Z}$ matrix
from the Gibbs sampler. Thus, we run two Markov chains simultaneously:
\begin{longlist}[(2)]
\item[(1)] A Gibbs sampler on the full model, to impute the missing data in
$\vef{Z}$ and estimate all parameters.
\item[(2)] A Metropolis--Hastings algorithm on $\delta$, in model space, to
find the best model. This algorithm uses an estimated Bayes factor
based on the current values in the Gibbs chain.
\end{longlist}
The aim is to search for $\delta^*$ such that $\delta^* = \arg\max
_{\delta} \mathrm{BF}_\delta$, but since we are not able to compute
$\mathrm{BF}_{\delta}$ exactly for any given $\delta$, we estimate
it using samples from the Gibbs sampler% of Section
, which yields a strongly consistent estimator. We then use the
estimated Bayes factor as the target in a stochastic search driven by a
Metropolis--Hastings algorithm.

A typical method of estimating a quantity such as (\ref{eqBFdef})
would be to use bridge sampling [\citet{MenWon96}]. However, since
the numerator and denominator have different dimensions (but the
numerator model is always nested in the denominator model), ordinary
bridge sampling will not work. A~variation [\citet{CheSha97}] which
accounts for this introduces a weight function to handle the dimension
difference. We summarize this strategy in the following proposition.
%
%pr1 #&#
%
\begin{prop}\label{propmain1}
$\!\!\!\!$Referring to (\ref{eqBFdef}), let $g(\veb{\theta})$
be such that $
\int p(\vef{Y}|\veb{\theta})g(\veb{\theta})\,
\mathrm{d} \veb{\theta}_{\delta^c} =
p_\delta(\vef{Y}|\veb{\theta}_\delta)
$. Then if expectation is taken with respect to the posterior distribution
$\pi(\veb{\theta} | \vef{Y})$,
\[
\mathbb{E} \biggl[\frac{\pi_\delta(\veb{\theta}_\delta) g(\veb
{\theta})}
{\pi(\veb{\theta})} \biggr] = \mathrm{BF}_{\delta}.
\]
\end{prop}

One particular $g$ function is defined as follows. Let $P_{\delta^c} :=
\vef{Z}_{\delta^c} (\vef{Z}^\prime_{\delta^c}
\vef{Z}_{\delta^c})^{-1} \vef{Z}^\prime_{\delta^c}$,
$\vef{C}_\delta:= (\vef{Y} - \vef{X} \veb{\beta} - \vef{Z}_{\delta}
\veb{\gamma}_{\delta})$, and
%
%e6 #&#
%
\begin{equation}\label{eqgfunction}
g(\veb{\theta}) = %\Pr(\ve{S}_{\delta^c} = \ve{s}_{\delta^c}) \times
(2\pi\sigma^2)^{-d^c/2} |\vef{Z}^\prime_{\delta^c}
\vef{Z}_{\delta^c}|^{1/2}
\times\exp\biggl(-\frac{1}{2\sigma^2}
\vef{C}_\delta^\prime
P_{\delta^c}
\vef{C}_\delta
\biggr),
\end{equation}
%
%Then if $\veb{\theta}^{(i)}$
%are samples from the posterior $\pi(\veb{\theta}|\vef{Y})$, the following
%estimator is strongly consistent for $\mathrm{BF}_\delta$.
which leads to the strongly consistent Bayes factor estimator
%
%e7 #&#
%
\begin{eqnarray}
\label{eqestimatorBF}
\widehat{\mathrm{BF}}_{\delta} &=& \frac{1}{N} \sum_{i=1}^N
\bigl( \phi^{2(i)}\bigr)^{d^c/2} \bigl|\vef{Z}^{(i)\prime}_{\delta^c}
\vef{Z}^{(i)}_{\delta^c}\bigr|^{1/2}\nonumber\\[-8pt]\\[-8pt]
&&\hphantom{\frac{1}{N} \sum_{i=1}^N}{}\times
\exp\biggl( -\frac{1}{2\sigma^{2(i)}} \biggl(
\frac{| \veb{\gamma}^{(i)}_{\delta^c} |^2}{\phi^{2(i)}} +
\vef{C}_{\delta}^{(i)\prime} P^{(i)}_{\delta^c}
\vef{C}_{\delta}^{(i)} \biggr)
\biggr).\nonumber
\end{eqnarray}
Details and proofs of the results given here are in
Supplemental Information, Section D [\citet{Lietal}].

%%%%%%%%%%%%%%%%%%%%%%%%%%%%%%%%%%%%%%%%%%%%%%%%%%%%%

%s3.3 #&#
\subsection{Increasing computational speed}\label{computationspeed}

For data sets with large numbers of SNPs and phenotypes, the slow
computation speed of the Gibbs sampler can be a major problem. We have
identified two bottlenecks. First, if the number of SNPs is increased,
then for each iteration, the number of missing SNPs to be
updated will also increase. Second, in the iterations of the Gibbs
sampler, the generation of
$\gamma$ involves inverting the matrix $\vef{Z}^\prime R^{-1}\vef
{Z}+(1/\phi^2)I$ each time, as the $\vef{Z}$ matrix changes at each
iteration. We address these in the following sections.

%s3.3.1 #&#
\subsubsection{SNP updating}
To speed up calculation, we show that instead of updating
all the SNPs at each iteration, updating only one column of SNPs (that
is, one SNP
updated for all observations) at each cycle will still conserve the
target stationary
distribution and ergodicity. As the SNP has only three possible values,
this change should not have a great effect on the mixing.

A consequence of this
result is that instead of updating tens or hundreds of SNPs in one cycle,
we need to update just one SNP in each cycle. This single-SNP updating
will dramatically speed up
computation, especially when there are large numbers of SNPs, or large
numbers of
observations, in the data. (See in
Supplemental Information [\citet{Lietal}], Section E.)

%s3.3.2 #&#
\subsubsection{Matrix inverse updating}

In the iterations of the Gibbs sampler, a~major bottleneck is the
generation of $\gamma$, since it involves inverting the matrix
$\vef{Z}^\prime R^{-1}\vef{Z}+(1/\phi^2)I$ each time, as the $\vef{Z}$
matrix changes at each iteration. Two modifications will speed up this
calculation, each based on Woodbury's formula [see \citet{Hag89}
and Appendix \ref{appwoodbury}].

By Woodbury's formula, if the matrices $A$ and
$I-VA^{-1}U$ are both invertible, then
%
%e8 #&#
%
\begin{equation}\label{eqwood1}
(A+UV)^{-1}= A^{-1}-A^{-1}U(I+VA^{-1}U)^{-1}VA^{-1}.
\end{equation}
If $U$ and $V$ are vectors, the inverse takes a particularly nice form:
%
%e9 #&#
%
\begin{equation}\label{eqwood2}
(A+u v^\prime)^{-1}=A^{-1}-\frac{A^{-1}u v^\prime A^{-1}}{(1+v^\prime
A^{-1}u)},
\end{equation}
so if we have $A^{-1}$, no further inversion is needed.

First, relating to the generation of $\gamma$ in (\ref{eqGibbs1}),
(\ref{eqwood1}) leads to the identity
%
%e10 #&#
%
\begin{equation}\label{eqeither}
\biggl(\vef{Z}^\prime R^{-1}\vef{Z}+\frac{1}{\phi^2}I\biggr)^{-1} =
\phi^2
\biggl[I-\vef{Z}^\prime
\biggl(\frac{1}{\phi^2}R+\vef{Z} \vef{Z}^\prime\biggr)^{-1}\vef
{Z}\biggr],
\end{equation}
where the left-hand side involves the inversion of an $s \times s $ matrix,
and the right-hand
side involves the inversion of an $n \times n$ matrix. Thus, we can
always choose to
invert the smaller matrix.

Next we look at inverting $\vef{Z}^\prime
R^{-1}\vef{Z}+(1/\phi^2)I$ [a similar argument can be developed for the
right-hand side of
(\ref{eqeither})]. Suppose, at the current iteration, we have
$A_0=\vef
{Z}_0^\prime R^{-1}\vef{Z}_0+(1/\phi_0^2)I$, and we
update to $A_1=\vef{Z}_1^\prime R^{-1}\vef{Z}_1+(1/\phi_1^2)I$. Because
we update one column of SNPs at each iteration, we have
$\vef{Z}_1=\vef{Z}_0 + \Delta$, where $\Delta$ is a
matrix of all $0$'s, except for one column. This column contains the
differences of
the respective columns from $\vef{Z}_1$ and $\vef{Z}_0$. Thus, $\Delta=
(0 \cdots0  0
\delta  0 \cdots0)$, and
\[
A_1 = A_0 + \Delta^\prime R^{-1} \vef{Z}_0 + \vef{Z}_0^\prime R^{-1}
\Delta+\Delta^\prime R^{-1} \Delta+\biggl(\frac{1}{\phi_1^2}
-\frac{1}{\phi_0^2}\biggr) I.
\]
The three matrices on the right-hand side involving $\Delta$ are all rank
one matrices, that is,
they are of the form $u v^\prime$ for column vectors $u$ and $v$.
Moreover, we can
write $I = \sum_{j=1}^s e_j e_j^\prime$, where $e_j$ is a column vector
of zeros with a
$1$ in the $j$th position. We can then apply (\ref{eqwood2})
three times to get the inverse of $A_1$. This calculation involves only
matrix by vector multiplications for the middle three terms on
the right-hand side. For the $e_j$ vectors, the multiplications reduce to an element
extraction. (See Appendix \ref{appwoodbury} for details.)
%%%%%%%%%%%%%%%%%%%%%%%%%%%%%%%%%%%%%%%%%%%%%%%%%%%%%

%%%%%%%%%%%%%%%%%%%%%%%%%%%%%%%%%%%%%%%%%%%%%%%%%%%%%
%s4 #&#
\section{Empirical analyses of BAMD}\label{secempirAnaly}

%s4.1 #&#
\subsection{Percentage of missingness handled}\label{subsecmissingness}
In this subsection we apply\break BAMD to simulated data in order to assess
the procedure's performance as we increase the percentage of missing
data in the $\vef{Z}$ matrix. We simulated a data set with six families, 20
observations in each family and 5 SNPs per observation. The five SNPs are
independent of each other. The six
families are also independent, so that the parents of the six families
are not
related and individuals across families are independent. On the
other hand, the individuals within each family share the same parents;
this relationship is captured via the numerator relationship matrix.
From this
data set, four data sets with different percentages of missing values,
5\%,
10\%, 15\%, and 20\%, were randomly derived. The family effects, $\beta$,
which were used to simulate the data, are listed in Table
\ref{tabbetaMissingness}. The true SNP effects (additive and dominant effects)
used to generate the data are listed in Table \ref{tabgammaMissingness}.
When simulating, we let the variance parameter $\sigma^2=1$. Our proposed
methodology was applied to analyze the data without missing values and
also to the new data containing missing values.

%t1 #&#
%
\begin{table}
\caption{The true
family effects for the simulated data set are given in the first row of the
table. The remaining rows indicate the estimated means returned from running
BAMD on the data sets derived by setting different degrees of missing values
in the SNP matrix of the simulated data set}
\label{tabbetaMissingness}
\begin{tabular*}{\tablewidth}{@{\extracolsep{\fill}}lcccccc@{}}
\hline
& $\bolds{\beta_1}$ & $\bolds{\beta_2}$ & $\bolds{\beta_3}$ &
$\bolds{\beta_4}$ & $\bolds{\beta_5}$ & $\bolds{\beta_6}$ \\
\hline
Actual value & 15\hphantom{.00} & 20\hphantom{.00} & 25\hphantom{.00}
& 30\hphantom{.00} & 35\hphantom{.00} & 40\hphantom{.00}
\\[4pt]
0\% missing & 15.45 & 20.65 & 25.48 & 29.84 & 34.76 & 40.40 \\
5\% missing & 15.16 & 20.74 & 25.46 & 28.29 & 33.43 & 38.62 \\
10\% missing & 16.18 & 21.38 & 25.65 & 30.71 & 35.86 & 40.81 \\
15\% missing & 15.45 & 19.63 & 24.59 & 30.18 & 35.38 & 40.18 \\
20\% missing & 14.87 & 20.18 & 24.68 & 30.08 & 34.88 & 40.13 \\
\hline
\end{tabular*}\vspace*{-2pt}
\end{table}

%t2 #&#
%
\begin{table}[b]
\tabcolsep=0pt
\caption{The true
additive and dominant effects for each SNP in the simulated data set
are given
in the first row of the table. The remaining rows indicate the estimated
SNP effects returned from running BAMD on the data sets derived by setting
different degrees of missing values in the SNP matrix of the simulated
data set}
\label{tabgammaMissingness}
{\fontsize{8.2pt}{11pt}\selectfont{
\begin{tabular*}{\tablewidth}{@{\extracolsep{\fill}}ld{2.2}ccd{2.2}cccd{2.2}d{2.2}c@{}}
\hline
& \multicolumn{1}{c}{\textbf{SNP1:a}} & \multicolumn{1}{c}{\textbf{SNP1:d}}
& \multicolumn{1}{c}{\textbf{SNP2:a}} & \multicolumn{1}{c}{\textbf{SNP2:d}}
& \multicolumn{1}{c}{\textbf{SNP3:a}} & \multicolumn{1}{c}{\textbf{SNP3:d}}
& \multicolumn{1}{c}{\textbf{SNP4:a}} & \multicolumn{1}{c}{\textbf{SNP4:d}}
& \multicolumn{1}{c}{\textbf{SNP5:a}} & \multicolumn{1}{c@{}}{\textbf{SNP5:d}} \\
\hline
Actual SNP & -2.00
& 1.00 & 1.00 & -1.00 & 3.00 & 0.00 & 2.50 & 0.10 & 0.30 & 3.00
\\[4pt]
0\% missing& -2.16 & 1.00 & 0.82 & -0.75 & 2.59 & 0.30 & 2.43 & 0.60 & -0.04
& 2.59 \\
5\% missing& -1.86 & 1.14 & 1.16 & -1.05 & 3.00 & 0.05 & 2.21 & -0.20 & 0.48
& 3.00 \\
10\% missing& -1.95 & 0.77 & 1.18 & -1.52 & 2.74 & 0.18 & 2.51 & 0.13 & 0.00
& 2.74 \\
15\% missing& -1.80 & 0.78 & 0.99 & -0.96 & 2.48 & 0.67 & 2.43 & 0.47 & 0.73
& 2.48 \\
20\% missing& -2.08 & 1.29 & 1.21 & -0.76 & 3.10 & 1.32 & 1.87 & -0.20
& 0.47
& 3.10 \\
%{\bf(0.077,0.142)}&{\bf(0.067,0.126)}&NC($3^\prime$UTR)\\
\hline
\end{tabular*}}}\vspace*{-3pt}
\end{table}

Note that for this small simulation, we used a parameterization
different from the $\{-1,0,1\}$ coding that we use for larger numbers
of SNPs.
In this example, each SNP effect is represented as $(\gamma_a, \gamma
_d)$---the additive and dominant effects of the SNP genotypes.

Tables \ref{tabbetaMissingness} and \ref{tabgammaMissingness}
summarize the parameter estimation capabilities of BAMD for family and
SNP effects. All calculations were based on samples obtained after an initial
burn-in of 20,000 iterations of BAMD. The results show that when the\vadjust{\goodbreak} percentage
of missing values is less than 15\%, the proposed methodology yields good
estimates for the parameters of direct interest. When the percentage
of missing values is greater than 15\%, we should be wary of interpreting
the results. For example, the true dominant effect for SNP 3 is
0, but the estimate is 1.32 when the percentage of missing values is
20\%.
Note that the estimate in this case is accurate
when the percentage of missing values is less than 10\%. We believe the
discrepancy arises because one category of genotype for SNP 3 has substantially
higher probability and it overpowers the other two categories. When the
percentage of
missing values increases, the dominated genotype category has only a small
chance to be well represented and thus may have unreliable estimates.

%t3 #&#
%
\begin{table}
\caption{The
true genotype probabilities for the SNPs used to generate the simulated
data set are given in the first 3 rows. The final row identifies the
frequency with
which the true genotype was imputed when running BAMD with 10\% of
missing data in the SNP matrix}
\label{tabfreqCorrectImputation}
{\fontsize{8.5pt}{11pt}\selectfont{
\begin{tabular*}{\tablewidth}{@{\extracolsep{\fill}}lccccc@{}}
\hline
& \textbf{SNP1} & \textbf{SNP2} & \textbf{SNP3} & \textbf{SNP4} & \textbf{SNP5} \\
& \multicolumn{1}{c}{$\bolds{a=-2}$}
& $\bolds{a=1}$ & $\bolds{a=3}$ & $\bolds{a=2.5}$
& $\bolds{a=0.3}$ \\
& $\bolds{d=1}$ & $\bolds{d=-1}$ & $\bolds{d=0.5}$
& $\bolds{d=0.1}$ & $\bolds{d=3}$\\
\hline
Actual SNP\\
\quad$\Pr(GG)$ & 0.1309 & 0.3012 & 0.8181 & 0.7719 & 0.3983 \\
\quad$\Pr(GC)$ & 0.5307 & 0.3875 & 0.0796 & 0.1950 & 0.5425 \\
\quad$\Pr(CC)$ & 0.3384 & 0.3113 & 0.1023 & 0.0331 & 0.0592 \\
[4pt]
Frequency of correct & 0.5500 & 0.5479 & 0.6337 & 0.8508 & 0.6516\\
imputation\\
%$\Pr(GG)$ & 0.0309 & 0.3012 & 0.3181 & {\bf0.3719} & 0.0983 \\
%$\Pr(GC)$ & { \bf0.8307} & 0.3875 & 0.0796 & 0.1950 & {\bf0.8425} \\
%$\Pr(CC)$ & 0.1384 & 0.3113 & {\bf0.6023} & 0.4331 & 0.0592 \\
%Frequency of correct imputation & {\bf0.7589} & 0.35052 & {\bf
%0.97621} &
%0.64829 & {\bf0.85301}\\
%%{\bf(0.077,0.142)}&{\bf(0.067,0.126)}&NC($3^\prime$UTR)\\
\hline
\end{tabular*}}}
\end{table}

Our ultimate goal is to identify significant SNPs from the candidate SNPs.
Since we believe that imputation is a tool to obtain better estimates
of the
parameters, we are not particularly interested in recovering the actual
imputed values for the missing SNPs. Nonetheless, the simulation
results in Table \ref{tabfreqCorrectImputation} show that when the
probability of one genotype for a certain SNP is dominantly high, the imputed
SNPs are correctly identified with probability ranging from $0.55$--$0.85$,
being correctly imputed more frequently than the other genotypes
distributions (see SNPs 3 and~4).
%The true probabilities of the SNPs are presented; those with genotypes
%that have a high probability are boldfaced. It is apparent that these
%are also imputed more frequently than the other possible genotypes for
%that SNP (see SNPs 3 and 4).

%%\subsection{Comparison with BIMBAM Through Simulations}

%s4.2 #&#
\subsection{Comparison with BIMBAM}
\label{subseccompare}

Here we compare our multiple-imputa\-tion missing data algorithm with a
program called BIMBAM [\citet{SerSte07}, \url
{http://stephenslab.uchicago.edu/software.html}],\break which is a~popular
program among geneticists for association genetics and variable
selection with missing data (using single imputation).

BAMD and BIMBAM both propose a two-stage procedure that involves
first finding a set of significant SNPs, and then running these
significant SNPs through a
variable selection procedure that finds the best subset of the significant
SNPs that describes the variation in the phenotype. Hence, in this
study, BIMBAM and BAMD are assessed through the SNPs they find in the first
stage and through the final model they put forth.

For the evaluation, we simulated data from the model given in
equation~(\ref{eqfixedeffects}).
The dimensions of the model were fixed to be $n=50$, $p=3$, and $s=25$
throughout. The three families comprised 16, 17, and 17 individuals, respectively.
In addition, the $\vef{X}$ and $\veb{\beta}$ matrices were
fixed. Entries in the $\vef{Z}$ matrix took three possible values,
mirroring the real-life situation, when they would represent genotypes.
Interest lies in discovering the significant coordinates of the
$\veb{\gamma}$ vector (which corresponds to SNP effects), in the
presence of
missing values in the $\vef{Z}$ matrix.

In the simulation study, three factors---percentage of missing values
in~$\vef{Z}$, magnitude of $\veb{\gamma}$ effects, and the degree of
correlation within a family---were varied across different levels. When
a particular factor was being investigated, the others were held
constant. Here in the main paper, we only present two specific
comparisons. The remainder of the results from the simulation study can
be found in Supplemental Information [\citet{Lietal}], Section~F. In\vadjust{\goodbreak}
running the study, we simulated several data sets for each case, and
observed very consistent results. Hence, in presenting our results, we
focus on a single representative data set in each case.

In both of the studies presented here, the $\veb{\gamma}$ vector was
generated from a~multivariate normal, and any values less than 3 in absolute
value were set to 0. After generating the $\vef{Y}$ responses, 20\% of the
entries in the $\vef{Z}$ matrix were set to missing before being passed to
BAMD and BIMBAM.

The first comparison measures the performance of the procedures when
an equicorrelation structure ($\rho$ was set to be 0.8)
exists within each of the three families. The second comparison presented here
aims to see if BAMD turns up many false positives. The $\veb{\gamma}$ vector
was generated in the same way as earlier, but only the coordinates with the
five largest values were kept. The rest were set to 0. In addition, the
individuals were assumed to be uncorrelated, that is, $\vef{R} = \vef{I}_n$.

%f1 #&#
%
\begin{figure}

\includegraphics{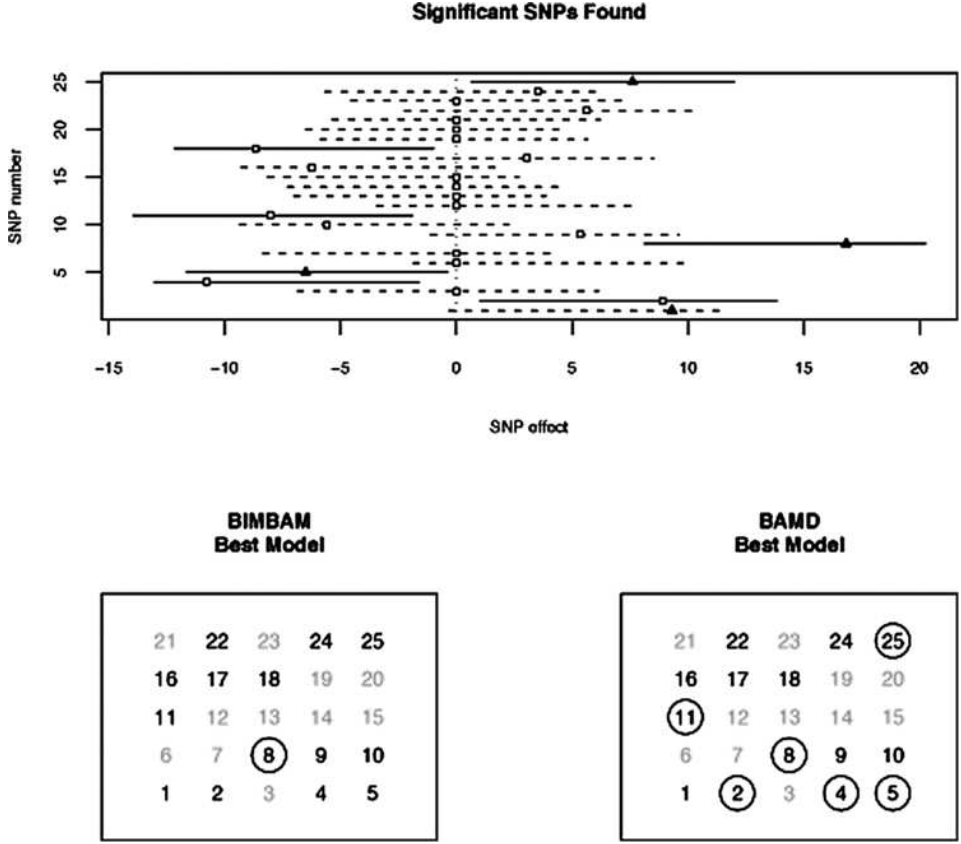}

\caption{In the upper panel,
triangles and squares represent the true
coordinates of the $\veb{\gamma}$ vector, where the true nonzero SNPs in
the model were
(1), (2), (4), (5), (8), (9), (10), (11), (16), (17), (18), (22), (24),
and (25). A solid triangle means that BIMBAM
found that SNP to be significant at $\alpha=0.05$ level, the remaining
SNPs are squares. Horizontal lines
represent highest posterior density intervals returned by BAMD.
Solid lines mean the 95\% HPD interval found that SNP to be
significant. Thus,
in the SNP-discovery stage, BIMBAM found SNPs (1), (5), (8), and (25) to
be significantly nonzero while BAMD picked out SNPs (2),
(4), (5), (8), (11), (18),
and (25). In the lower panel, the gray numbers are SNPs that were
exactly 0 in the true model, and the black numbers are SNPs with
nonzero effects. The circled numbers are the SNPs
that were in the best model found by the procedure. Thus, the best
model found
by BIMBAM contains only SNP (8), whereas the best model found by BAMD contains
SNPs (2), (4), (5), (8), (11), and (25).}
\label{figeqCorr}
\end{figure}

\begin{figure}

\includegraphics{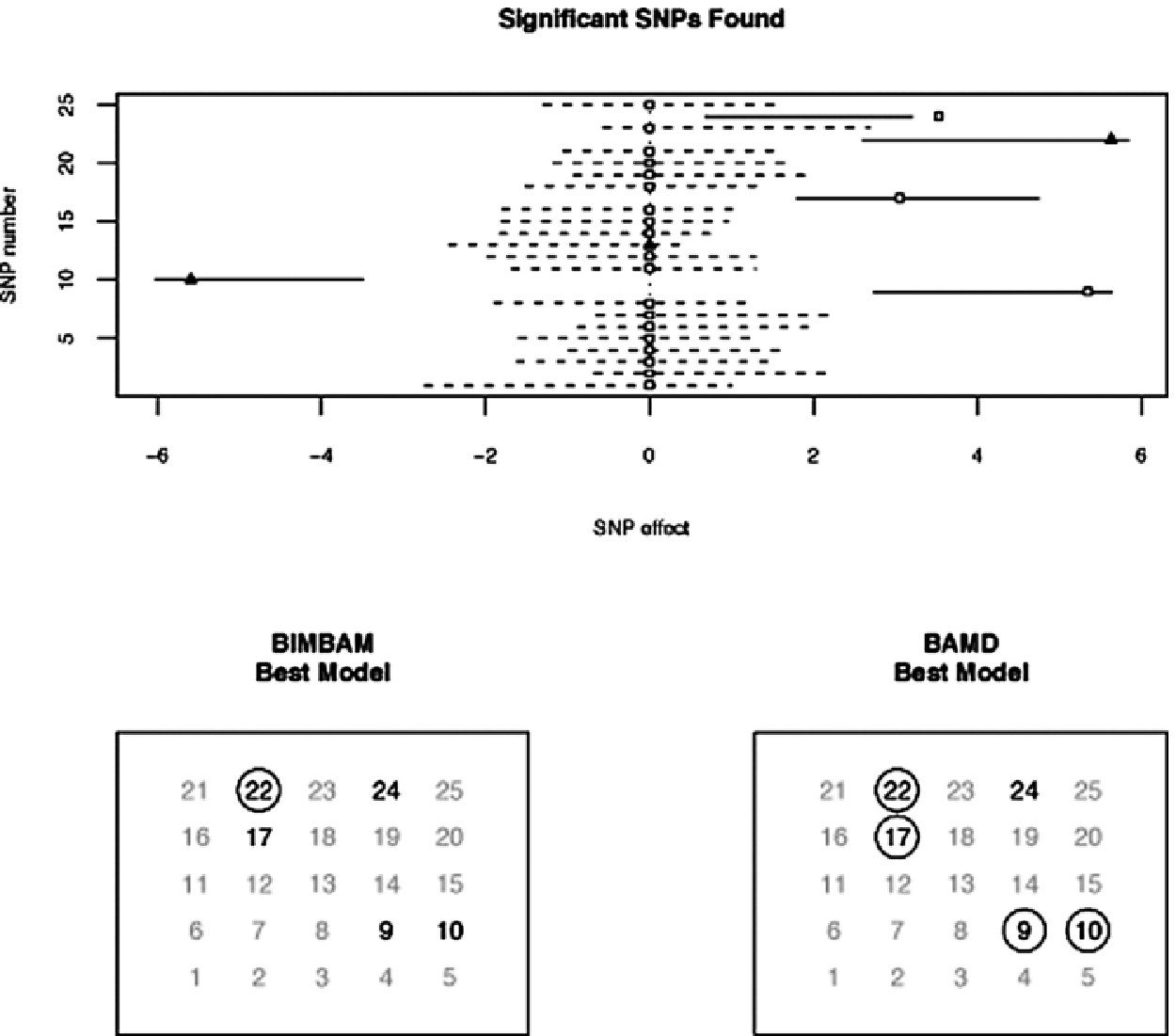}

\caption{In the upper panel triangles and
squares represent the true
coordinates of the $\veb{\gamma}$ vector, where the true nonzero SNPs in
the model were (9), (10), (17), (22), and (24). A~solid triangle means
that BIMBAM
found that SNP to be significant at $\alpha=0.05$ level, the remaining
SNPs are squares. Horizontal lines
represent highest posterior density intervals returned by BAMD.
Solid lines mean the 95\% HPD interval found that SNP to be
significant. Thus,
in the SNP-discovery stage, BIMBAM found SNPs (10), (13), and (22) to be
significantly nonzero while BAMD found SNPs (9), (10), (17), (22), and (24).
In the lower panel, the gray numbers are SNPs that were
exactly 0 in the true model, and the black numbers are SNPs with
nonzero effects. The circled numbers are the SNPs
that were in the best model found by the procedure.
The best model found by BIMBAM contains only SNP (22), whereas the best model
found by BAMD contains SNPs (9), (10), (17), and~(22).}
\label{figfp}
\end{figure}

The results for each comparison are summarized through two diagrams.
The first (the upper panels in Figures \ref{figeqCorr} and
\ref{figfp}) display the SNPs that
BAMD and \mbox{BIMBAM} found to be significant in the first stage.
The lower panels in Figures~\ref{figeqCorr} and \ref{figfp}
display the output from the variable selection procedure.

%f2 #&#
%

Each figure shows that BAMD significantly outperforms BIMBAM. In the
first example, BIMBAM found only one of the $14$ significant SNPs,
while BAMD found six. In the second example, there were five significant
SNPs, and BIMBAM only found one again, while BAMD found four of
them.\vspace*{-3pt}

%s5 #&#
\section{Analysis of the loblolly pine data}
\label{secanalysis}

%Here we took data from a recently published paper as an example
%to illustrate the power of our association model.
Carbon
isotope discrimination (CID) is a time-integrated trait measure
of water use efficiency. \citet{Gonetal08}
used the family-based approach of the Quantitative Transmission
Disequilibrium Test QTDT to detect SNPs associated with CID. We
utilized the family structure of this population [also
described in \citet{Kayetal05}] in the design matrix in
our model. Of the 61 control-pollinated families measured for
CID, each had approximately 15 offspring that were clonally
propagated by rooted cuttings to generate the ramets (genetically
identical replicates). Each
genotype has two ramets, sampled from each of two
testing sites at Cuthbert, GA and Palatka, FL. Our approach
enables us to utilize the family pedigree and parental
information to recover missing SNP genotypes. With informative
priors, we infer the progeny SNP genotype through Mendelian
randomization [\citet{FalMac}]. With uninformative
priors, we assume SNPs are missing at random and assign equal
probability for each genotype class for missing SNPs.

All SNPs are simultaneously tested under our association model.
The Gibbs sampler ran for 50,000 iterations. The first 10,000
iterations were burn-in, after which we thinned the chain every
4 iterations; the autocorrelation reduced significantly after
thinning (data not shown). Thus, we have a total of 10,000
samples for each chain for our statistics, and we then applied the
variable selector
on the four~SNPs that were found when using the informative prior.\vadjust{\goodbreak}

%t4 #&#
%
\begin{table}
\caption{Significant SNPs from QTDT
tests and the results
from the BAMD association model, with~95\%~confidence intervals}
\label{tabBAMD}
\begin{tabular*}{\tablewidth}{@{\extracolsep{\fill}}lccc@{}}
\hline
& \multicolumn{3}{c@{}}{\textbf{SNP}}\\[-4pt]
& \multicolumn{3}{c@{}}{\hrulefill}\\
& \multicolumn{1}{c}{\textbf{Informative prior}}
& \multicolumn{1}{c}{\textbf{Uninformative prior}} &
\\
& \textbf{95\% C.I.} & \textbf{95\% C.I.} &
\multicolumn{1}{c@{}}{\textbf{Type}$\bolds{^\ddagger}$} \\
\hline
(3) caf1\_s1$^*$&($-$0.008, 0.110)\hphantom{$-$}&\textbf{(0.013, 0.129)}&Syn\\
(5) ccoaomt\_s10$^{*\dagger}$&
\textbf{($\bolds{-}$0.103, $\bolds{-}$0.012)}& \textbf{($\bolds{-}$0.097,
$\bolds{-}$0.005)}&NC(intron)\\
(6) cpk3\_s5&\textbf{($\bolds{-}$0.052, $\bolds{-}$0.004)}
&($-$0.048, 0.001)\hphantom{$-$}&Syn\\
(29) dhn1\_s2$^{*\dagger}$&
\textbf{(0.065, 0.113)}&(0.044, 0.092)&NC($3^\prime$UTR)\\
(31) ein2\_s1$^{*\dagger}$&
\textbf{(0.077, 0.142)}&\textbf{(0.067, 0.126)}&NC($3^\prime$UTR)\\
\hline
\end{tabular*}
\legend{$^*$ Indicates significant in Gonz{\'a}lez-Mart{\'i}nez et~al.
(\citeyear{Gonetal08}). Bold type indicates significant at the 5\%
level from our association testing, the rest being nonsignificant.
$^\dagger$ indicates presence in best model found by variable selector.
As indicated in Gonz{\'a}lez-Mart{\'i}nez et~al.
(\citeyear{Gonetal08}), there are additional SNPs that are marginally
significant at $\alpha=0.1$, which we also detected. $^\ddagger$: Syn,
synonymous SNP; NC, noncoding; UTR, untranslated region.}
\end{table}

We detected significant effects of several SNPs on CID at a 95\%
Bayesian confidence interval (Table \ref{tabBAMD}). Using the
uninformative prior, we found 3 significant SNPs [(3)
ccoaomt\_s10, (5) ein2\_s1, (31) Caf1\_s1]. Using the
informative prior, we detected 4 SNPs [(5) ein2\_s1, (6)
cpk3\_s5, (29) dhn1\_s2, (31) Caf1\_s1] as significant. Note
that (6) and (29) are close to significant using the uninformative
prior, and (3) was close to significant using the informative
prior. This suggests that for these data, the effect of the
prior information is important. The QTDT
test resulted in 4 significant SNPs, (3), (5), (29), (31), all
of which were detected by BAMD which, in addition, found SNP
(6). Moreover, it is important to note that BAMD detected
these SNPs simultaneously, an indication that their collective
effect on the phenotype is being detected.

The use of tag SNPs in a pedigree does not allow for ``fine mapping''
to SNP effects [\citet{NeaIng08}, \citet{autokey7}]. Thus,
the effects of these SNPs on carbon isotope discrimination may reflect
the involvement of many linked genes on the phenotype.

\begin{figure}

\includegraphics{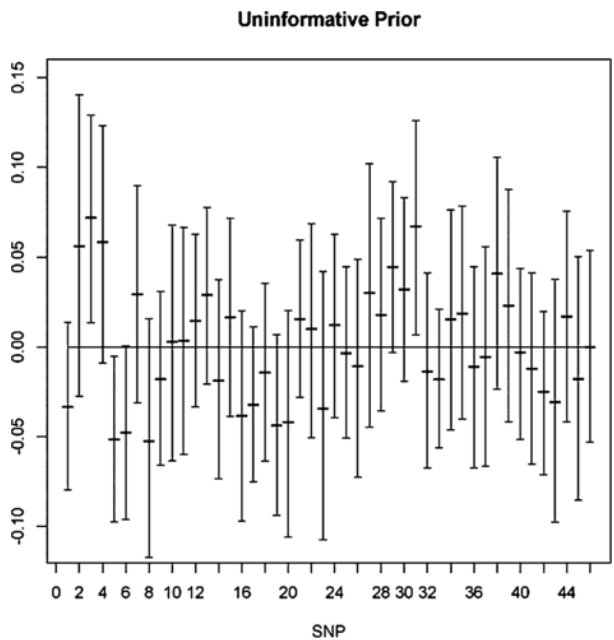}

\caption{$95\%$ Confidence intervals
for the $44$ SNPs
from the carbon isotope data, based on 10,000 Gibbs samples from the BAMD
model using uninformative priors (equal probability) for the missing SNPs.
The significant SNPs are those with intervals that do not cross $0$,
SNPs (3)
\textit{caf1}, (5) ccoaomt, and (31) \textit{ein2}.}
\label{figuninfo}
\end{figure}

\begin{figure}

\includegraphics{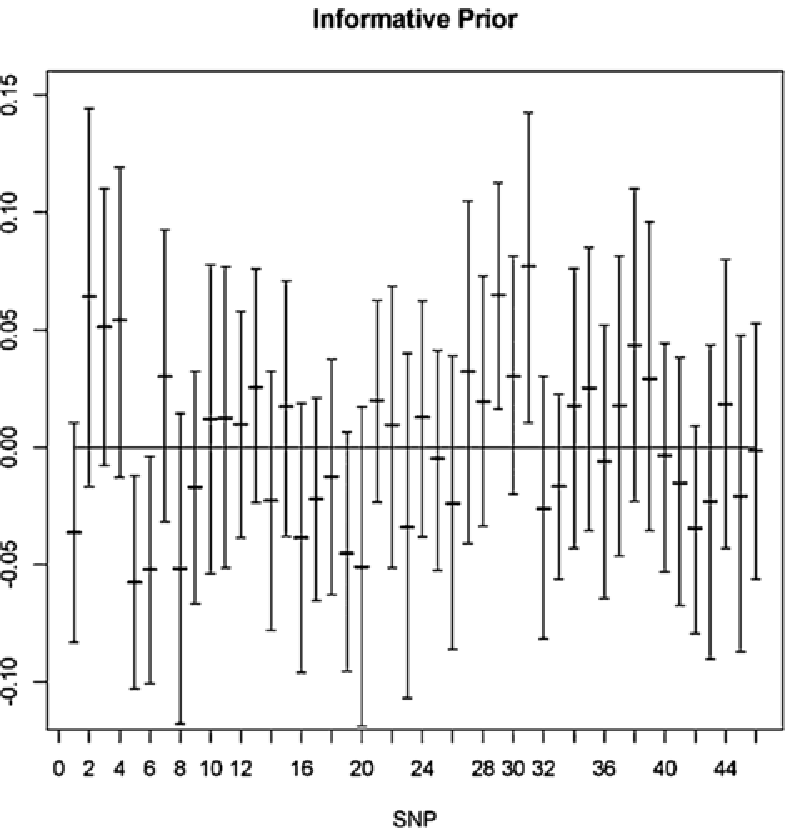}

\caption{$95\%$ Confidence intervals
for the $44$ SNPs from the carbon
isotope data, based on 10,000 Gibbs samples from the BAMD model using
informative priors (Mendelian randomization) for the missing SNPs. The
significant SNPs are those with intervals that do not cross $0$,
SNPs (5) ccoaomt, (6) \textit{cpk3}, (29) \textit{dhn1}, and (31)
\textit{ein2}.}
\label{figinfo}
\end{figure}

We also provide Figures \ref{figuninfo} and \ref{figinfo}, showing
the results for all of the SNPs in the data set. Figure \ref
{figuninfo} is based on using uninformative SNP priors, while Figure
\ref{figinfo} uses informative priors. Although there are few
differences in the graphs (showing the strength of the data with
respect to the model), we see that the prior can matter. For example,
SNP (3) is significant when the noninformative prior is used, but not
so when we use the informative prior. The opposite finding holds for
SNP (6). Looking at the figures, we see that the significant intervals
only barely cross zero; thus, the inclusion of relevant prior
information can be quite important.

%f3 #&#
%
%

%f4 #&#
%

The four SNPs picked out when using the informative prior were (5), (6),
(29), and
(31). Due to the small number of variables under consideration, the variable
selector procedure was able to run through all 16 possible models.
The one with the highest Bayes factor was found to contain SNPs (5),
(29), and
(31).

%s6 #&#
\section{Discussion}
\label{secdisc}

Association testing is being applied to discover relationships among
SNPs and complex
traits in plants and animals [\citet{autokey7},
\citet{HirDal05}, \citet{Bal06}, \citet{Zhuetal08}].
Our model was developed specifically for detecting associations in
loblolly pine data, but can be applied to other species as well. Here
we discuss some features and limitations of the method.
%In this paper we
%propose a Bayesian hierarchical approach to association testing
%with missing data.

\subsection*{Multiple imputation} Multiple imputation of missing SNP
data is
the best way to ensure unbiased parameter estimates, which is
an important consideration given that SNP effects tend to be
small for complex traits of greatest biological interest, and
given that results of association studies typically motivate
more detailed and labor-intensive investigations of how and why
associations were detected.

We used simulation to compare BAMD and BIMBAM for their detection of
``correct'' vs. ``incorrect'' SNPs, and found that BAMD performed better
than \mbox{BIMBAM}. In practice, this advantage of BAMD over BIMBAM would
likely be greatest when missing SNPs are not in LD with nearby SNPs (or
adjacency cannot be determined). This is the case in many species,
including loblolly pine, in which LD is low and genomic resources such
as high-resolution genomic maps and high-density SNP chips for genome
scanning are not as well developed as they are for the human genome.
The higher computational intensity required for formal multiple
imputation in BAMD is a trade-off, however, this has not restricted its
practical use in most data sets. For very large data sets, parallel
processing seems a logical next step in further increasing the
computational efficiency of BAMD.

\subsection*{Family structure} Our method can be applied to
family-based association populations, populations of unrelated
genotypes, or
combination populations. It can incorporate prior information
if known. [The application of BAMD in \citet{QueGopCum} was to a
population of unrelated genotypes, where significant SNPs related to
disease resistance were found.]
%Moreover, the method can be easily adapted to discrete phenotypes
%using a
%probit link, by adding a latent variable in the Gibbs sampler.

\subsection*{Probit models} Although here we assume a continuous response
variable, the
method can be adapted to discrete phenotypes using a
probit link. For example, in a case control study, the response
would be either case or control status, and with a probit model
we add a latent variable in the Gibbs sampler. %This
%extension can also be modified to handle the situation of a
%multiple category response.

\subsection*{SNP detection}
Although BAMD successfully detected the same
significant SNPs as were previously detected using the family-based
method QTDT [\citet{Gonetal08}], as well as an
additional significant SNP, the BAMD variable selector indicated that a
subset of the significant SNPs was sufficient to explain variation in
the phenotype carbon isotope discrimination. This is a useful tool for
biologists because a simultaneous solution for SNP effects enables
detection of numerous SNPs that collectively explain phenotypes, which
in turn enables further biological experiments to investigate their
underlying basis.

However, the candidate SNPs found by BAMD and QTDT cannot necessarily
be deemed ``correct'' or ``incorrect'' without additional biological
experiments. As such, little more can be stated about the correctness
of SNPs 3, 5, 29, and 31 without validation experiments. In the
broader context of association testing, it is relevant to note that the
use of QTDT is limited to families, whereas BAMD and BIMBAM can be used
to detect associations in families as well as populations of unrelated
individuals. The ability to use BAMD and BIMBAM in many different
types of populations is appealing.

\subsection*{Simultaneous vs. genome-wide} Genome-wide association
studies are, typically, marginal processors of the data. That is, each
SNP is assessed individually for its association, so simultaneous
assessment, or epistasis, cannot be detected. A~model such as (\ref
{eqfixedeffects}) is assessing the SNPs simultaneously---that is its
strength. But how many SNPs should we expect to be able to handle in
one model? Computational issues aside, if the number of SNPs is greatly
increased, we are then susceptible to the usual regression
pitfalls---multicollinearity being the most prevalent. Thus, we
recommend using BAMD on smaller sets of SNPs that have had some
preprocessing. Thus far, BAMD has been used successfully on a model
with $400$ SNPs [\citet{QueGopCum}], and we have tested it on as
many as $800$ SNPs.

\subsection*{Missing data}
The missing data problem is common across all genomics data sets, so
there is broad potential utility of this method. The assumption of MAR
(missing at random), which is reasonable in these contexts, may bear
additional research attention. If there are quality concerns about SNP
data, there are some statistical steps forward, as noted by
\citet{Wiletal10}, such as using indicator variables of
missingness as predictors. This approach can even be extended to test
if missingness is a heritable trait and, if so, the MAR assumption is
invalid. Next generation sequencing platforms may generate sufficient
data to enable this assumption to be tested and, if borne out, may
motivate placement of priors on SNP calls in certain sequence contexts.

Last, software to run the Gibbs sampler and variable selector is
available in the {\tt R} package BAMD.

\begin{appendix}\label{app}
%s7 #&#
\section{Calculating the numerator relationship~matrix}
\label{appnumerator}

The algorithm is due to \citet{autokey12} and \citet{Quaas}.
The individuals within $61$ families and the parents for the $61$
families are ordered together\vadjust{\goodbreak} such that the first $1,\ldots,a$ subjects
are unrelated and are used as a ``base'' population. Let the total
number of subjects within families and parents of the $61$ families
be~$n$, and we will get a numerator relationship matrix with dimension
$n \times n$. As the first $a$ subjects (being part of the parents of
the $61$ families ) are unrelated, the upper left submatrix with
dimension $a\times a$ of the numerator relationship matrix is identity
matrix $I$. This identity submatrix will be expanded iteratively until
it reaches to dimension $n\times n$.

As we know the sub-numerator relationship matrix for the first
unrelated~$a$ subjects is the identity, next we will give the
details how to calculate the remaining cells of the numerator
relationship matrix for the related subjects. Consider the
$j$th and the $i$th subject from the above ordered subjects:
\begin{longlist}[(2)]
\item[(1)] If both parents of the $j$th individual
are known, say, $g$ and $h$, then
\begin{eqnarray*}
R_{ji}&=&R_{ij}=0.5(R_{ig}+R_{ih}),\qquad i=1,\ldots, j-1;
\\
R_{jj}&=&1+0.5 R_{gh},
\end{eqnarray*}
where $R_{ji}$ is the cell of the numerator
relationship matrix in the $j$th row and $i$th column.
\item[(2)] If
only one parent is known for the $j$th subject, say, it is $g$,
then
\begin{eqnarray*}
R_{ji}&=&R_{ij}=0.5 R_{ig},\qquad i=1,\ldots,j-1;
\\
R_{jj}&=&1.
\end{eqnarray*}
\item[(3)] If neither parent is known for the $j$th subject,
\begin{eqnarray*}
R_{ji}&=&R_{ij}=0,\qquad i=1,\ldots,j-1;
\\
R_{jj}&=&1.
\end{eqnarray*}
\end{longlist}
For the loblolly pine data, we have $44$ pines acting as
grandparents and they produce $61$ pine families. The $61$
families contains $888$ individual pine trees all together,
also called clones. The phenotypic responses are taken from
the individual clones. So our interest is in calculating the
relationship matrix for the $888$ clones and it would have a
dimension $888\times888$. According to Henderson's method,
we ordered the $44$ grandparent pines and $888$ individual
pines together such that the first $a$ pines are not related.
Starting from the $(a+1)$th pine, we applied the above
iteration calculation algorithm, and in the end had a
relationship matrix with dimension $932\times932$ for all the
grandparent pines and all individual clones. We took a
submatrix from the right bottom of the previous numerator
relationship matrix with dimension $888\times888$ and it is
the numerator relationship matrix we used in the loblolly pine
data analysis.

%s11 #&#
\section{Estimation with the EM algorithm}\label{appEM}
%s11.1 #&#
\subsection{Missing data}
The EM algorithm begins by building the \textit{complete data
likelihood}, which is the likelihood function that would be used
\textit{if the missing data were observed}. When we fill in the missing
data we write\vadjust{\goodbreak} $Z_i^* = (Z_i^o, Z_i^m)$, and the \textit{complete data} are
$(Y, Z^*)$ with likelihood function
%
%e22 #&#
%
\begin{eqnarray}\label{eqCDLike}
L_C &\propto& \prod_{i \in I_0}\exp\biggl( - \frac{1}{2 \sigma
^2}(Y_i-X_i\beta-Z_i \gamma)^2\biggr) \nonumber\\[-8pt]\\[-8pt]
&&{}\times\prod_{i \in I_M}\exp\biggl( -
\frac{1}{2 \sigma^2}(Y_i-X_i\beta-Z_i^* \gamma)^2\biggr),\nonumber
\end{eqnarray}
where $I_o$ indexes those individuals with complete SNP data, and $I_M$
indexes those individuals with missing SNP information.

The \textit{observed data likelihood}, which is the function that we
eventually use to estimate the parameters, must be summed over all
possible values of the missing data. So we have
\begin{eqnarray*}
L_o &\propto&\frac{1}{(2\pi\sigma^2)^{n/2}}\prod_{i \in I_0}\exp
\biggl(- \frac{1}{2 \sigma^2}(Y_i-X_i\beta-Z_i \gamma)^2\biggr)\\
&&{}\times\prod_{i \in I_M} \sum_{Z_i^*}\exp\biggl( - \frac{1}{2 \sigma
^2}(Y_i-X_i\beta-Z_i^* \gamma)^2\biggr).\nonumber
\end{eqnarray*}
The distribution of the missing data $Z_i^*$ is given by the ratio of $L_C/L_o$:
%
%e23 #&#
%
\begin{equation}\label{eqmissingdatadist}
P(Z_i^*)=\frac{\exp( - (Y_i-X_i\beta-Z_i^*
\gamma)^2/({2 \sigma^2}))}{\sum_{Z_i^*}\exp( -
(Y_i-X_i\beta-Z_i^* \gamma)^2/({2 \sigma
^2}))},
\end{equation}
where the sum in the denominator is over all possible realizations of
$Z_i^*$. This is a discrete distribution on the missing SNP data for
each individual. To understand it, look at one individual.

Suppose that there are $g$ possible genotypes (typically $g=2$ or $3$)
and individual $i$ has missing data on $k$ SNPs. So the data for
individual $i$ is $Z_i=(Z^o, Z^m)$, where $Z^m$ has $k$ elements, each
of which could be one of $g$ classes. For example, if $g=3$ and $k=7$,
then $Z^m$ can take values in the following:\vspace*{4pt}
\begin{center}
\begin{tabular}{c|c|c|c|c|c|c|c|}
\multicolumn{1}{c}{}&\multicolumn{7}{c}{SNP}\\
\cline{2-8}
&$\ast$&&&&$\ast$&&\\
\cline{2-8}
Genotype&&$\ast$&$\ast$&&&$\ast$&\\
\cline{2-8}
&&&&$\ast$&&&$\ast$\\
\cline{2-8}
\end{tabular}
\end{center}

\vspace*{6pt}

\noindent where the $\ast$ show one possible value of the $Z_i^m$. For the
example, there are $3^7= 2187$ possible values for $Z_i^m$. In a real
data set this could grow out of hand. For example, if there were $12$
missing SNPs, then there are 531,441 possible values for $Z_i^m$; with
20 missing SNPs the number grows to 3,486,784,401 (3.5 billion).

%s11.2 #&#
\subsection{An EM algorithm}
To the expected value of the log of the complete data likelihood (\ref
{eqCDLike}), we only deal with the second term (with the missing\vadjust{\goodbreak}
data). This expected value does not change the piece with no missing
data, but does change the second piece. Standard calculations give
\[
\mathrm{E} \biggl(\frac{1}{2 \sigma^2}(Y_i-X_i\beta-Z_i^* \gamma
)^2\biggr)
= \frac{1}{2 \sigma^2}\bigl(Y_i-X_i\beta-\mathrm{E}(Z_i^*) \gamma\bigr)^2 +
\operatorname{Var}(Z_i^* \gamma),
\]
where
%
%e24 #&#
%
\begin{equation}\label{eqexpectations}
\mathrm{E}(Z_i^*) = (Z_i^o, \mathrm{E}(Z_i^m)) \quad\mbox{and}\quad
\operatorname{Var}(Z_i^*\gamma) = \operatorname{Var}(Z_i^m\gamma_i^m).
\end{equation}
If we define
\[
Y_{n \times1} = \pmatrix{
Y_1\cr Y_2\cr
\vdots\cr
Y_n},\qquad
X_{n \times p} = \pmatrix{
X_1\cr X_2\cr \vdots\cr X_n},\qquad
\vef{Z}_{n \times s} = \pmatrix{
(Z_1^o, \mathrm{E}(Z_1^m))\vspace*{2pt}\cr (Z_2^o, \mathrm{E}(Z_2^m))\cr
\vdots
\cr(Z_n^o, \mathrm{E}(Z_n^m))},
\]
the expected complete data log likelihood is
%
%e25 #&#
%
\begin{equation}\label{eqExplogCDLike}
\mathrm{E} \log L_C = -\frac{n}{2} \log\sigma^2 - \frac{1}{2 \sigma
^2} |
Y-X\beta-\vef{Z} \gamma|^2 -\frac{1}{2 \sigma^2} \gamma^\prime V_Z
\gamma,
\end{equation}
where $V_{Z_i}$ is the variance--covariance matrix of the vector $Z_i$
with elements given by
\[
V_{{Z_i}_{jj^\prime}} =\cases{
0, &\quad if either $Z_{ij}$ or $Z_{ij^\prime}$ is
observed, \vspace*{2pt}\cr
\operatorname{Cov}(Z_{ij},Z_{ij^\prime}), &\quad if neither $Z_{ij}$
nor $Z_{ij^\prime}$ is observed,}
\]
and $V_Z = \sum_{i \in I_M}V_{Z_i}$.
Standard calculus will show that the MLEs from (\ref{eqExplogCDLike})
are given by
%
%e26 #&#
%
\begin{eqnarray}\label{eqMLEs}
\hat\beta&=& (X^\prime X)^{-1} X^\prime(Y-\vef{Z}\hat\gamma
),\nonumber\\
\hat\gamma&=& (\vef{Z}^\prime\vef{Z}-V_Z)^{-1} \vef{Z}^\prime(I-H)Y,
\\
\hat\sigma^2 &=& \frac{1}{n} ( | Y-X \hat\beta-\vef{Z} \hat
\gamma
|^2 + \hat\gamma^\prime V_Z \hat\gamma)\nonumber.
\end{eqnarray}
The algorithm now iterates between (\ref{eqmissingdatadist}), (\ref
{eqexpectations}), and (\ref{eqMLEs}) until convergence.

%s11.3 #&#
\subsection{Implementation}
To implement the EM algorithm, we must be able to either:
\begin{longlist}[(1)]
\item[(1)] calculate the expectation and variance in (\ref
{eqexpectations}), or
\item[(2)] generate a random sample from (\ref{eqmissingdatadist}) and
calculate the terms in (\ref{eqexpectations}) by simulation.
\end{longlist}
The first option is impossible and the second is computationally
intensive, but the only way.

Going back to (\ref{eqmissingdatadist}), note that this is the
distribution of the vector of missing values for individual $i$. If the
data are $Z_i^* = (Z_i^o, Z_i^m)$, we are only\vadjust{\goodbreak} concerned with $Z_i^m =
(Z^m_{i1}, \ldots, Z^m_{ik})$, and for $\mathbf{c} = (c_1,\ldots,c_k)$,
\[
P(Z_i^m= \mathbf{c}_0 )=\frac{\exp( - (Y_i-X_i\beta-Z_i^o \gamma_i^o-
\mathbf{c}_0\gamma_i^m)^2/({2 \sigma
^2})
)}{\sum
_{\mathrm{all}\ \mathbf{c}}\exp( - (Y_i-X_i\beta
-Z_i^o \gamma_i^o- \mathbf{c}\gamma_i^m)^2/({2 \sigma
^2}))},
\]
where the sum in the denominator can easily have over $1$ billion terms.

A possible alternative is to use a Gibbs sampler to simulate the
distribution of $Z_i^m$ by calculating the distribution of each element
conditional on the rest of the vector. For a particular element
$Z^m_{ij}$, the conditional distribution given the rest of the vector
$Z^m_{i(-j)}$ is given in (\ref{eqconditional}). %
So to produce a sample of~$Z_i^m$, we loop through a Gibbs sampler.

Unfortunately, there may be problems with this algorithm in that it may
still be too computationally intensive. The Gibbs samplers (\ref
{eqGibbs1}) and (\ref{eqconditional}) need to be run for every
iteration of the EM algorithm. For each iteration of EM we may need
20--50 thousand Gibbs iterations. If there is a lot of missing data,
this could result in a very slow algorithm.

%s10 #&#
\section{Matrix inverse updates}\label{appwoodbury}
We are interested in matrices of the form $A_0 + \sum_{k=1}^p u_k
v_k^\prime$, where
$u_k, v_k$, $k=1, \ldots, p$, are vectors. For this form we have the
following lemma,
which follows from Woodbury's formula.
%
%le1 #&#
%
\begin{lemma}
Let $A_0$ be invertible, and $u_j, v_j$, $j=1, \ldots, p$, be vectors. Define
\[
A_k = A_0 + \sum_{j=1}^k u_j v_j^\prime.
\]
Then for $k=1, \ldots, p$,
%
%e21 #&#
%
\begin{equation}\label{eqrecursion}
A_k^{-1} = A_{k-1}^{-1} - \frac{A_{k-1}^{-1}u_{k} v^\prime_k
A_{k-1}^{-1}}{1+v^\prime_k
A_{k-1}^{-1}u_k}.
\end{equation}
\end{lemma}

Then, to calculate $A_p^{-1} = ( A_0 + \sum_{k=1}^p u_k
v_k^\prime
)^{-1}$,
we can start with $A_0^{-1}$, and use the recursion to get to
$A_p^{-1}$. Note that
each step of the recursion requires only multiplication of matrices by vectors.
Moreover, in many applications the vectors $u_k, v_k$ are sparse, so
the multiplication
amounts to extracting elements.
\end{appendix}

\begin{supplement}[id=suppA]
\stitle{Theory and additional simulations}
\slink[doi]{10.1214/11-AOAS516SUPP} %[doi,text={...}] - jei reikia
%suskaldyti doi
\slink[url]{http://lib.stat.cmu.edu/aoas/516/supplement.pdf}
\sdatatype{.pdf}
\sdescription{The Supplemental Information contains details on the variable selector, and
the proof of convergence of the two Markov chains (the Gibbs sampler and the
model search).  In addition, there are further comparisons between BAMD and
BIMBAM.}
\end{supplement}

% imsref loaded by lrinkeviciute, 2012-01-13 08:44:15
% imsref loaded by lrinkeviciute, 2012-01-13 08:54:42

\printaddresses

\end{document}